\documentclass[aps,prl,twocolumn,superscriptaddress]{revtex4}
\usepackage{graphicx}
\usepackage{amssymb}
\usepackage{amsmath}
\usepackage{graphicx}
\usepackage{epsfig}
\usepackage{color}
\usepackage{bm}

\begin{document}

\title{Nonlocal Kondo effect and quantum critical phase in heavy fermion metals}
\author{Jiangfan Wang}
\affiliation{Beijing National Laboratory for Condensed Matter Physics, Institute of Physics,
Chinese Academy of Science, Beijing 100190, China}
\author{Yi-feng Yang}
\email[]{yifeng@iphy.ac.cn}
\affiliation{Beijing National Laboratory for Condensed Matter Physics,  Institute of Physics, 
Chinese Academy of Science, Beijing 100190, China}
\affiliation{School of Physical Sciences, University of Chinese Academy of Sciences, Beijing 100190, China}
\affiliation{Songshan Lake Materials Laboratory, Dongguan, Guangdong 523808, China}
\date{\today}

\begin{abstract}
Heavy fermion metals typically exhibit unconventional quantum critical point or quantum critical phase at zero temperature due to competition of Kondo effect and magnetism. Previous theories were often based on certain local type of assumptions and a fully consistent explanation of experiments has not been achieved. Here we develop an efficient algorithm for the Schwinger boson approach to explore the effect of spatial correlations on the Kondo lattice and introduce the concept of nonlocal Kondo effect in the quantum critical region with deconfined spinons. We predict a global phase diagram containing a non-Fermi liquid quantum critical phase with a hidden holon Fermi surface and a partially enlarged electron Fermi surface for strong quantum fluctuations while a single quantum critical point for weak quantum fluctuations. This explains the unusual metallic spin liquid  recently reported in the frustrated Kondo lattice CePdAl and resolves the Fermi volume puzzle in YbRh$_2$Si$_2$. Our theory highlights the importance of nonlocal physics and provides a unified understanding of heavy fermion quantum criticality.
\end{abstract}

\maketitle

The interplay of antiferromagnetic (AFM) transition and $f$ electron delocalization underlies many exotic properties of the Kondo lattice physics \cite{YangPNAS2017,review_si2010}. In particular, the recent discovery of a non-Fermi liquid (NFL) quantum critical phase in the frustrated Kondo lattice CePdAl has posed an urgent challenge in clarifying the nature of this intermediate state \cite{zhao2019CePdAl}, which is in stark contrast with the usual observation of a single quantum critical point (QCP) in many heavy fermion antiferromagnets, such as YbRh$_2$Si$_2$ \cite{Custers2003Nature_YRS,Paschen2004-YRS-Hall,friedemann2010fermi_YRS,prochaska2020singular_YRS}, CeRhIn$_5$ \cite{Park_CeRhIn5_2006,CeRhIn5dHvA}, and CeCu$_{6-x}$Au$_{x}$ \cite{Lohneysen1998-CeCuAu,Schroder2000CeCuAu,Lohneysen2008-CeCuAu,Lohneysen2009-CeCuAu}. In the latter case, the AFM QCP is often thought to be accompanied with the full delocalization of $f$-electrons into a heavy Fermi liquid (HFL), possibly manifested by an abrupt change of the electron Fermi surface from ``small" (no $f$ electrons) to ``large" (with $f$ electrons). In CePdAl, however, the two transitions are detached. The intermediate phase spans over a broad range of the pressure-magnetic field phase diagram and is neither magnetically ordered nor a Fermi liquid. Similar intermediate phase has been observed previously in Ir or Ge-doped YbRh$_2$Si$_2$ \cite{friedemann2009detaching,Custers2010prl_Ge-YRS}. Its origin is unclear, but often attributed to magnetic frustrations, low dimensionality, or large spin/orbital degeneracy \cite{si2006global,coleman2010frustration,Pixley2014quantum,tomita2015YbAlB}.

The lack of a thorough microscopic understanding lies in the extreme difficulty of simulating the Kondo lattice. The widely used dynamical mean-field theory \cite{Georges1996RMP} and its cluster extensions \cite{martin2008clusterDMFT,kotliar2011cdmft} can well capture local or short-range correlations but fail to describe long-range quantum critical fluctuations. Exact lattice simulations often require heavy computational efforts and can only be applied under very special conditions on small lattices \cite{Sato2018MC}. In this regard, the recent development of the large-$N$ Schwinger boson approach represented an important advance \cite{Coleman-SWB-2006,lebanon2006conserving,lebanon2007fermi,Yashar1D,Komijani2019,wang2019quantum}. Compared to the prevalent slave-boson method, the Schwinger boson representation of spins allows for a better treatment of local moment antiferromagnetism and its interplay with Kondo screening. However, its latest implementations on the Kondo lattice have all predicted direct transitions between AFM and HFL, showing no sign of an intermediate phase \cite{Komijani2019,wang2019quantum}.

The discrepancy comes from the local approximation adopted in these calculations, which ignores momentum dependence of quasiparticle self-energies in order to reduce  the computational efforts \cite{Komijani2019,wang2019quantum}. To overcome this issue, we go beyond the local approximation and develop an efficient numerical algorithm to solve the Schwinger boson self-consistent equations with full frequency and momentum-dependent self-energies. This enables us to study the low-energy charge and spin dynamics with both temporal and spatial fluctuations. Our method is then applied to the two-dimensional (2D) Kondo-Heisenberg model on the square lattice and finds in certain parameter range an emergent intermediate state featured with gapless spinon and holon excitations, and a partially enlarged (or ``medium") electron Fermi surface due to the generalized Luttinger sum rule \cite{coleman2005sum, pepin2005fractionalization}, which is forbidden in the local approximation. The phase diagram and finite temperature properties are controlled by the interplay of a deconfined AFM QCP and a transition to the ``large" electron Fermi surface, which merge together into a single transition for large spin size. The key of our finding is a nonlocal Kondo effect mediated by holons propagating on the lattice. Our results explain the recent experiments in CePdAl and YbRh$_2$Si$_2$ and provide a unified theory of heavy fermion quantum criticality.

We start with the following Hamiltonian:
\begin{eqnarray}
H=t\sum_{\left\langle ij\right\rangle}c_{i\alpha a}^\dagger c_{j\alpha a}+J_K\sum_{i}{\bf S}_{i}\cdot {\bf s}_{i}+J_H\sum_{\left\langle ij\right\rangle}{\bf S}_{i}\cdot {\bf S}_{j},
\label{H}
\end{eqnarray}
where $c_{i\alpha a}^\dagger $ creates a conduction electron of spin $\alpha$ and channel (orbital) $a = 1,2,\cdots,K$ on site $i$, ${\bf s}_{i}$ is its spin operator, and ${\bf S}_{i}$ denotes the local spin. The Schwinger boson approach enlarges the SU(2) spin group to the symplectic group Sp($N$) such that ${\bf S}_{i}\rightarrow S_{i}^{\alpha \beta}=b_{i\alpha}^\dagger b_{i\beta}-\tilde{\alpha}\tilde{\beta}b_{i,-\beta}^\dagger b_{i,-\alpha}$, where $b_{i\alpha}$ represents the Schwinger boson (spinon), $\alpha=\pm 1,\cdots,\pm N/2$, and $\tilde{\alpha}=\text{sgn}(\alpha)$ \cite{FlintSpN2009}. A local constraint is then imposed to reduce the enlarged Hilbert space to physical subspace, $n_{b,i}\equiv \sum_{\alpha}b_{i\alpha}^\dagger b_{i\alpha}=2S$, which may be implemented by introducing the Lagrange multiplier, $\sum_{i} \lambda_i(n_{b,i}-2S)$. A biquadratic exchange term, $-\zeta J_H\sum_{\left\langle ij\right\rangle}\left( {\bf S}_{i}\cdot {\bf S}_{j} \right)^2$, is often included to avoid artificial first-order transitions at large $N$, which can be absorbed into the quadratic term under SU(2) symmetry \cite{fczhang2002pathology}. Depending on the ratio of $2S/K$, there exist three distinct regions, where the local spins are either underscreened ($2S/K>1$), overscreened ($2S/K<1$), or exactly-screened ($2S/K=1$) \cite{parcollet1997transition}. We focus on the exactly-screened case. The Kondo and Heisenberg terms can be factorized using two auxiliary fields:
\begin{eqnarray}
\frac{J_K}{N} S_i^{\alpha\beta}c_{i\beta a}^\dagger c_{i\alpha a} \rightarrow \frac{1}{\sqrt{N}}b_{i\alpha}^\dagger c_{i\alpha a}\chi_{ia}+h.c.+\frac{|\chi_{ia}|^2}{J_K}, \notag \\
\frac{J_H}{N} S_i^{\alpha\beta}S_j^{\beta\alpha} \rightarrow  \tilde{\alpha}b_{j,-\alpha}^\dagger b_{i,\alpha}^\dagger \Delta_{ij}+h.c.+\frac{N|\Delta_{ij}|^2}{J_H},
\end{eqnarray}
where $\Delta_{ij}$ denotes the spin-singlet valence bond on adjacent sites and $\chi_{ia}^\dagger$ can be viewed as a composite fermion of the Kondo state formed by a conduction hole and a spinon. $\chi_{ia}^\dagger$ is also called the holon field since it carries a positive electric charge and has no spin.

\begin{figure}
\centering\includegraphics[scale=0.167]{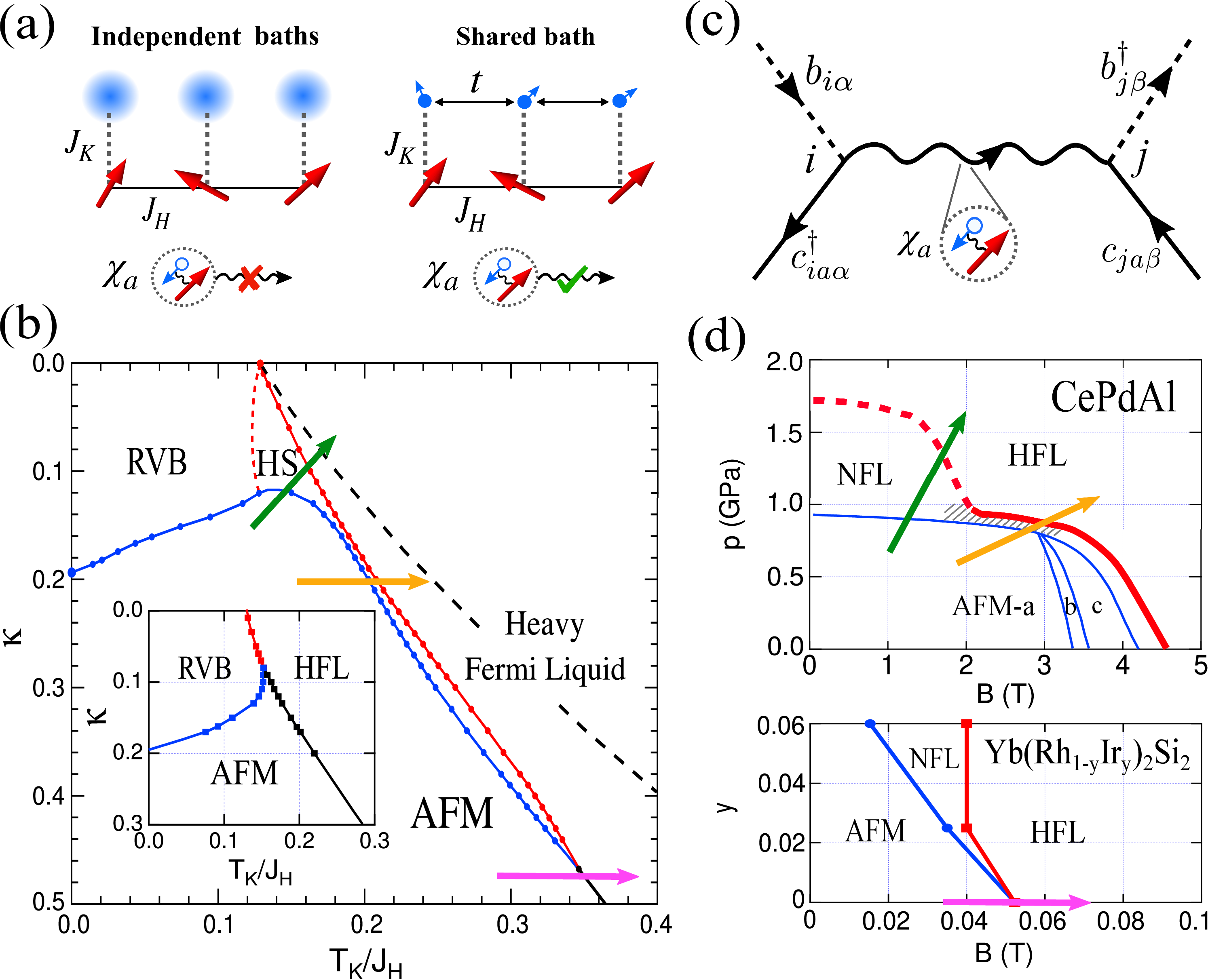}
\caption{(a) Illustration of the Kondo-Heisenberg model with independent electron baths where holons are not allowed to propagate on the lattice (left) and a shared bath in this work (right). (b) Theoretical phase diagram in the large $N$ limit on the $\kappa$ and $T_K/J_H$ plane, showing four phases: the N\'eel state (AFM), the resonating valence bond (RVB) state, the heavy Fermi liquid  (HFL), and the intermediate holon state (HS). The RVB state may turn into the valence bond solid (VBS) at finite $N$. The AFM and HFL phase boundaries are separated for small $\kappa$ but merge together for $\kappa \geq 0.47$. The intermediate HS region is missing under local assumption (inset). The dashed line inside the HFL marks a transition from $\Delta\neq 0$ (short-range magnetic correlations) to $\Delta=0$ (a local Fermi liquid). The arrows indicate different routes of quantum phase transitions. (c) The Feynman diagram of nonlocal Kondo scattering mediated by propagating holons.  (d) Experimental phase diagrams of CePdAl \cite{zhao2019CePdAl} and Yb(Rh$_{1-y}$Ir$_y$)$_2$Si$_2$ \cite{friedemann2009detaching}. The shadow in the phase diagram of CePdAl marks the region with linear-in-$T$ resistivity. The arrows mark possible correspondences with those in (b).}
\label{fig:Global}
\end{figure}

To proceed, we assume the mean-field variables $\lambda_i=\lambda$ and $\Delta_{i,i+\hat{x}}=\Delta_{i,i+\hat{y}}=\Delta$. The former replaces the local constraint $n_{b,i}=2S$ by the average spinon occupation and the latter describes a candidate spin liquid energetically favored in the Heisenberg model \cite{Read1991largeSpN}. The rotational symmetry is preserved under combined operation of lattice rotation and gauge transformation \cite{Wen2002quantum}. In the large-$N$ limit, the spinon and holon self-energies are \citep{supp}:
\begin{eqnarray}
\Sigma _{b}({\textbf p}, i\nu_n)&=&-\frac{\kappa}{\beta\mathcal{V}} \sum_{{\textbf k}m}g_c({\textbf p}-{\textbf k}, i\nu_n-i\omega_m ) G_{\chi }({\textbf k}, i\omega_m ), \notag \\
\Sigma _{\chi }({\textbf p}, i\omega_m)&=&\frac{1}{\beta\mathcal{V}}\sum_{{\textbf k}n}g_c({\textbf k}-{\textbf p}, i\nu_n -i\omega_m ) G_{b}({\textbf k}, i\nu_n ),
\label{eq:SelfE}
\end{eqnarray}
where $g_c$ is the bare Green's function of conduction electrons, $G_{b}$ and $G_{\chi }$ are the full Green's functions of spinons and holons to be self-consistently determined by their self-energies, $\omega_m$ ($\nu_n$) are the fermionic (bosonic) Matsubara frequencies, $\beta$ is the inverse temperature, and $\mathcal{V}$ is the total number of lattice sites. The parameter $\kappa\equiv 2S/N=K/N$ controls the effective strength of quantum fluctuations. The self-energy of conduction electrons is absent in the large-$N$ limit, thus preventing proper treatment of electric transport. In previous calculations \cite{Yashar1D,Komijani2019,wang2019quantum}, a local approximation was adopted to reduce the computational efforts by ignoring the momentum dependence of the self-energies. This is equivalent to assign independent electron baths for each local spin as illustrated in Fig.~\ref{fig:Global}(a). Under this approximation, only direct phase transitions are allowed as shown in the inset of Fig.~\ref{fig:Global}(b). To overcome this issue, we notice that the momentum convolution can be turned into a simple multiplication in the coordinate space, $\Sigma_{b/\chi}({\textbf r})\sim g_c({\textbf r})G_{\chi/b}({\textbf r})$, which motivates us to develop an efficient algorithm based on the fast Fourier transform (FFT) and solve the above equations in coordinate space without approximation \cite{supp}.

Figure \ref{fig:Global}(b) plots the resulting zero-temperature phase diagram on the $\kappa$ and $T_K/J_H$ plane, where $T_K=De^{-2D/J_K}$ is the single ion Kondo temperature and $D$ is the half bandwidth of conduction electrons. $T_K/J_H$ is also called the Doniach ratio. The phase diagram contains four regions: the AFM N\'eel order,  the resonating valence bond (RVB) state with a small electron Fermi surface, the HFL with a large electron Fermi surface, and the intermediate holon state (HS) with gapless spinon and holon excitations. Our result is in close resemblance to the experimental phase diagram of CePdAl (Fig. \ref{fig:Global}(d)), showing different (narrow or wide) regions of intermediate NFL phase tuned by pressure and magnetic field \cite{zhao2019CePdAl}. The intermediate phase disappears for $\kappa> 0.47$ with weak quantum fluctuations, where the AFM and HFL transitions merge together to give a single quantum critical point as in YbRh$_2$Si$_2$ tuned by Ir doping \cite{friedemann2009detaching}. Inside the HFL, short-range magnetic correlations may vanish ($\Delta=0$) at large $T_K/J_H$ and we enter a local Fermi liquid with independently screened spins.

The RVB, AFM and HFL phases are already present in the local approximation and can be largely understood by two limits. In the Heisenberg limit ($J_K=0$), the spins are decoupled from conduction electrons. At finite $N$, the RVB state will turn into the valence bond solid (VBS) state due to spinon confinement with the inclusion of monopoles  \cite{read1989valence,read1990spinpeierls}. The AFM N\'eel order is associated with spinon condensation. The transition between them marks a deconfined QCP with divergent spinon confinement length, where monopoles are irrelevant \cite{senthil2004deconfined}. In this particular model, the VBS-AFM and RVB-AFM transitions are described by the same critical theory with gapless spinons and emergent U(1) gauge fields \cite{Senthil2005}. Our calculations reproduce the scaling of the staggered susceptibility $\chi_{st}\propto T e^{4\pi \rho_s/T}$ ($\rho_s$: the spin stiffness) in the renormalized classical regime above the N\'eel order and $\chi_{st}\propto T^{-2+\eta}$ in the quantum critical regime \cite{CHN1989,Chubukov1994}, with the anomalous dimension $\eta$ approaching unity at the critical $\kappa$, reflecting deconfined free spinons \cite{Kaul2008}. In the limit of $\kappa\rightarrow 0$, both spinons and holons are localized and the lattice physics is reduced to a collection of decoupled spins ($\Delta=0$) undergoing independent Kondo screening beyond a critical $T_K/J_H$. For finite $\kappa$, the local approximation predicts direct transitions between three states, supporting local quantum criticality \cite{si2001localQCP}.

\begin{figure}
\centering\includegraphics[scale=0.46]{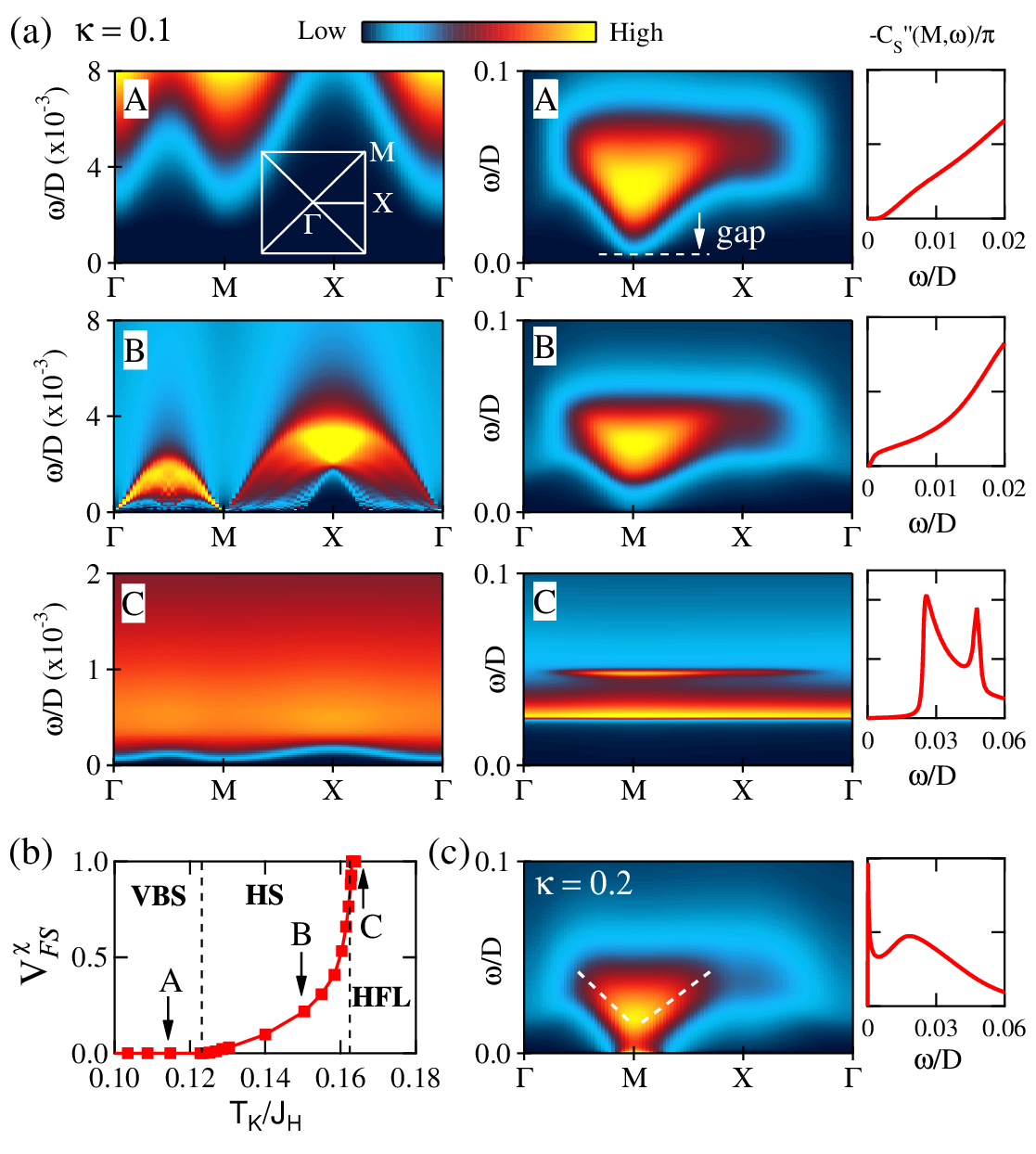}
\caption{(a) Holon (left panel) and spin (right panel) excitation spectra along the high symmetry line of the Brillouin zone (inset) at $\kappa=0.1$ for $T_K/J_H=0.115$, $0.15$ and $0.164$ (from top to bottom) with different ground states as marked in (b). The color represents the intensity of the spectral functions $-C_{n_\chi}''({\textbf k},\omega)/\pi$ and $-C_{S}''({\textbf k},\omega)/\pi$, respectively. The spin spectra at $M$ are also plotted for clarity. (b) Evolution of the holon Fermi volume, $V_{\text{FS}}^\chi$, as a function of $T_K/J_H$ at $\kappa=0.1$. (c) The spin spectral function at $\kappa=0.2$ and $T_K/J_H=0.2$ at a small but finite $T$ right above the AFM ground state. The white dashed lines are a guide to the eye.}
\label{fig:HS}
\end{figure}

By contrast, our calculations with momentum dependent quasiparticle self-energies reveal an intermediate phase (HS) with gapless spinon and holon excitations for $\kappa<0.47$. Importantly, we obtain the correct zero temperature AFM instability for both zero and finite $J_K$, in conformity with the Mermin-Wagner theorem \cite{MW}, while the local approximation predicted falsely a finite temperature transition for nonzero $J_K$ \citep{wang2019quantum}. Lattice propagations are crucial for both results, which yield a dispersive holon band and a holon Fermi surface determined by the poles of $G_\chi$ at the Fermi energy or, equivalently, the effective Kondo coupling, $J_K^*({\textbf p})\equiv [J_K^{-1}+\text{Re}\Sigma_\chi({\textbf p},0)]^{-1}$. Physically, the momentum dependence of $J_K^*({\textbf p})$ implies an unusual nonlocal and cooperative scattering process described by $J_K^*(\bm{r}_j-\bm{r}_i)c_{ja\beta}^\dagger b_{j\beta} b_{i\alpha}^\dagger c_{ia\alpha}$, in which a conduction hole and a spinon form a spinless quasi-bound state (holon) at $\bm{r}_i$, propagate to another site $\bm{r}_j$, and then unbound themselves (see Fig. \ref{fig:Global}(c)). Such ``nonlocal Kondo effect" mediated by fractional quasiparticles underlies the emergent NFL state (HS) between AFM and HFL and differs conceptually from earlier proposal of partial Kondo screening where local spins and electrons are both intact \cite{Pixley2014quantum,Sato2018MC}. The presence of a holon Fermi surface may help further stabilize the deconfinement by coupling to the U(1) gauge field at finite $N$ and making monopole fluctuations irrelevant \cite{Hermele2004,Lee2008}, while the gauge field is either ``Higgsed" or confined in other three phases, forbidding fractional excitations at low temperatures.

The holon Fermi volume, $V_{\text{FS}}^\chi=\mathcal{V}^{-1}\sum_{{\textbf p}}\theta(-J_K^*({\textbf p}))$, is gauge invariant and evolves continuously in the intermediate state, as plotted in Fig.~\ref{fig:HS}(b) for $\kappa=0.1$. It satisfies the generalized Luttinger sum rule, $N V_{\text{FS}}^c-V_{\text{FS}}^\chi=n_c$, where $V_{\text{FS}}^c$ is the Fermi volume of conduction electrons and $n_c$ is the electron number per channel (orbital) \citep{coleman2005sum}. The sum rule reflects the electric charge conservation associated with the global U(1)  symmetry: $\chi_{ia}\rightarrow \chi_{ia}e^{i\phi_a}$, $c_{i\alpha a}\rightarrow c_{i\alpha a}e^{-i\phi_a}$. As a result, the Fermi surface of conduction electrons is ``small" ($NV_{\text{FS}}^c=n_c$) in the RVB phase, ``large" ($NV_{\text{FS}}^c=n_c+1$) in the HFL, but ``partially" enlarged (or ``medium") in between, consistent with the calculated electron Fermi surface with $1/N$ correction \cite{supp}. For the local approximation, holons have no dispersion and their Fermi volume is either zero or unity, thus preventing a partially enlarged electron Fermi surface. We note that the HS phase is different from the FL$^*$ phase discussed in Ref. \cite{Senthil2003prl}. The latter is characterized by decoupled spinons and a small electron Fermi surface, similar to our RVB state. It is also different from a conventional two-band metal with no fractional excitations but only (dispersive) intact $f$-electrons \cite{Senthil2004weak}.

More detailed information on the low-energy spin and charge excitations in the intermediate state can be extracted from the holon density-density correlation function $C_{n_\chi}=-\frac{1}{K}\left\langle  n_\chi(\bm{r}_i,\tau) n_\chi(\bm{r}_j,\tau')\right\rangle_c$, where $n_\chi(\bm{r}_i,\tau)=\sum_a|\chi_{ia}(\tau)|^2$, and the dynamic spin structure factor $C_S=-\frac{1}{N}\left\langle S_i^z(\tau) S_j^z(\tau')\right\rangle_c$ with $S_{i}^z=\sum_\alpha\tilde{\alpha}b_{i\alpha}^\dagger b_{i\alpha}$. The subscript ``$c$'' denotes that only connected diagrams are considered. Figure~\ref{fig:HS}(a) plots their imaginary parts in the energy-momentum space at $\kappa=0.1$ for specially chosen values of $T_K/J_H$. We find that both excitations are gapped in the RVB (panel A) and HFL (panel C) phases. In the intermediate state (panel B), the holons become gapless around $\Gamma$ and ${\text M}$, corresponding to particle-hole pairs from same or different parts of the holon Fermi surface. With increasing $T_K/J_H$, the holon bands ($D_\chi$) become increasingly narrow, implying a heavy effective mass as large as $m^*_\chi/m_e\propto D/D_\chi\approx 10^4$ near the HFL boundary. Across the boundary, the holon Fermi surface vanishes in the HFL state. Accordingly, conduction electrons achieve a large Fermi surface following the Luttinger sum rule. The spin excitation spectra in the intermediate state are also gapless but highly damped. For the square lattice model in the Heisenberg limit, the spectra are sharply defined but gapped outside of the AFM phase. Here the coupling with holons smears out the gap and results in gapless but damped spin excitations. For comparison, Fig.~\ref{fig:HS}(c) shows the results for $\kappa=0.2$ at small but finite temperature above the AFM ground state. We find a sharp peak near zero energy as the precursor of spinon condensation. This distinguishes the spinon dynamics  inside the intermediate phase. At $\kappa=0.48$, the AFM QCP marks a direct transition to the HFL and is featured with both critical spinons and heavy holons \cite{supp}. Given the gapless charge and spin excitations, the Kondo screening may exist in a critical way inside the HS. Its difference from the fully Kondo screened HFL may also be reflected in two-particle correlation functions \cite{Chalupa2021Fingerprints}, which unfortunately require two-loop diagrams beyond our numerical capability.

Thus, the phase diagram for $\kappa>0.1$ is largely controlled by the interplay of a deconfined AFM QCP and a transition to the large electron Fermi surface. At finite temperature, one may further expect a crossover line connecting to the renormalized classical regime in the Heisenberg limit above the AFM order, and a delocalization line associated with the transition to the ``large" electron Fermi surface. In between, irrespective of an intermediate state or a QCP at zero temperature, there always exists a paramagnetic region with short-lived spinon and holon excitations and a ``partially" enlarged electron Fermi surface. This provides a candidate microscopic interpretation of the two fluid model with coexistent spin liquid and heavy quasiparticles \cite{Yang2012,Yang2016}. The partially enlarged electron Fermi surface evolves with temperature, supported lately by angle-resolved photoemission spectroscopy (ARPES) \cite{Chen2017CeCoIn5} and ultrafast optical pump-probe spectroscopy \cite{Liu2020pump} in CeCoIn$_5$. The fact that it varies continuously and is ``nearly large" in the vicinity of the single QCP for large $\kappa$ might help resolve the recent controversy in YbRh$_2$Si$_2$, where, contrary to usual expectation based on the change of the Hall coefficient \cite{Paschen2004-YRS-Hall}, ARPES reported a large Fermi surface above the AFM (70 mK) \cite{YRSarpes2015}. Even deep on the AFM side, ``band bending" has been observed in paramagnetic CeRhIn$_5$ \cite{Chen2018CeRhIn5}. The crossover in the Hall coefficient might be explained if holon contributions are taken into consideration \cite{zhang2018CePdAl,Nair2012,Coleman2001}. The presence of deconfined holons is a peculiar feature of the intermediate state in our theory, whose consequences are yet to be fully elaborated. In the AFM phase at zero temperature, a small electron Fermi surface is always expected due to spinon condensation.

\begin{figure}
\centering\includegraphics[scale=0.45]{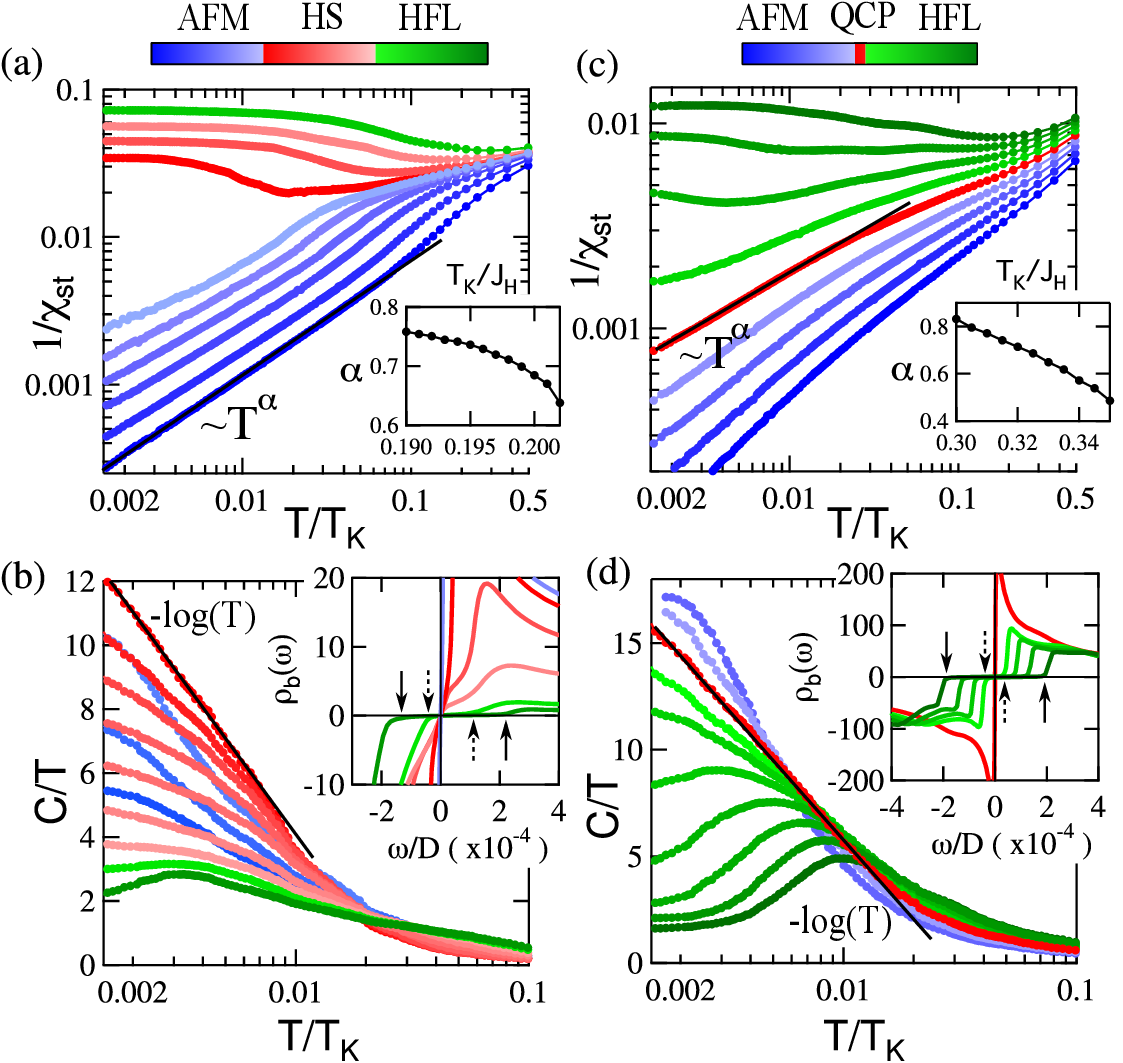}
\caption{Temperature dependence of (a) the inverse staggered susceptibility and (b) the specific heat coefficient at $\kappa=0.2$ for different values of $T_K/J_H$. The colors distinguish the AFM (blue), HS (red), and HFL (green) regions. The inset of (a) shows the power-law exponent $\alpha$ as a function of $T_K/J_H$ on the AFM side, and that of (b) compares the low temperature spinon density of states in the HS and HFL regions. The arrows mark the gap edges. (c) and (d) are similar plots at $\kappa=0.48$, where the red color denotes the QCP. }
\label{fig:phy2}
\end{figure}

Some of the physical properties in the intermediate region can be approximately captured by the large-$N$ limit. Figure \ref{fig:phy2} plots the calculated staggered magnetic susceptibility $\chi_{st}$ and specific heat coefficient $C/T$ at $\kappa=0.2$ and $0.48$. In both cases, we see $\chi_{st}\sim T^{-\alpha}$ on the AFM side and $C/T\sim -\ln T$ at the AFM QCP, typical of NFL. The exponent $\alpha$ varies monotonically with $T_K/J_H$ and drops rapidly near the AFM QCP (roughly 0.5 at $\kappa=0.48$). Its value is much smaller than that of the Heisenberg model, reflecting the presence of additional holon excitations. Its nonuniversality seems consistent with experimental observations, where $\alpha$ varies from $1/3$ in UCu$_{5-x}$Pd$_x$ \cite{AronsonUCuPd1996} to $0.51$ in Ce(Ru$_{1-x}$Fe$_x$)$_2$Ge$_2$ \cite{CeRuFeGe2003} to 0.75 in CeCu$_{5.9}$Au$_{0.1}$ \cite{Schroder2000CeCuAu}, apart from other possible reasons. Inside the HFL, $C/T$ shows a broad maximum at finite temperature. This difference may be understood from the insets of Figs.~\ref{fig:phy2}(b) and \ref{fig:phy2}(d), where the spinon density of states is singular at the AFM QCP but gapped in the HFL. Inside the intermediate state, the spinon density of states is gapless but nonsingular and $C/T$ keeps growing with lowering temperature, reflecting nonuniversal NFL behaviors as observed in CePdAl \cite{zhao2019CePdAl}.

It remains to be seen how transport properties might be affected when the electron self-energy is included at finite $N$. A linear-in-$T$ resistivity has been proposed for critical holons \cite{Komijani2019} or due to spinon scattering with vanishing holon velocity \cite{pepin2005fractionalization}. This might be true near the AFM QCP, but must not be extended to the whole intermediate region where spinon and holon dynamics are not always critical, even for frustrated Kondo lattices. Indeed, the linear-in-$T$ resistivity was only observed over a narrow region of the intermediate phase in CePdAl \cite{zhao2019CePdAl}. Its appearance is probably associated with the distorted Kagome structure, calculations on which require more auxiliary fields and will be left for future work.

The intermediate phase has also been observed in other compounds including YbRh$_2$Si$_2$ with Ir or Ge doping \cite{friedemann2009detaching,Custers2010prl_Ge-YRS} and YbAgGe under field \cite{Canfield2011YbAgGe}. These compounds adopt different crystal structures, suggesting that it is not a phenomenon solely for frustrated Kondo lattices. Large spin/orbital degeneracy and low dimensionality may also introduce strong quantum fluctuations \cite{coqblinschrieffer1969}. Of course, details of the phase diagram may be altered by finite $N$ corrections including gauge fluctuations. Nevertheless, our approach allows for the possibility of the intermediate state, which is an advance beyond the local approximation. Key features distinguishing our theory from the conventional Hertz-Millis theory include the partially enlarged electron Fermi surface with an intermediate Fermi wave vector \cite{coleman2005sum}, the existence of multiple charge carriers, and possibly singular charge fluctuations \cite{Komijani2019,prochaska2020singular_YRS}. More elaborate studies along this line may lead to a better understanding of the Kondo lattice physics. 

This work was supported by the National Key R\&D Program of China (Grant No. 2017YFA0303103), the National Natural Science Foundation of China (Grants No. 12174429, No. 11774401, No. 11974397), and the Strategic Priority Research Program of the Chinese Academy of Sciences (Grant No. XDB33010100).

\end{document}

\title{Nonlocal Kondo effect and quantum critical phase in heavy fermion metals\\
\vspace{0.2cm}
- Supplemental Material -}
\author{Jiangfan Wang}
\affiliation{Beijing National Laboratory for Condensed Matter Physics, Institute of Physics,
Chinese Academy of Science, Beijing 100190, China}
\author{Yi-feng Yang}
\email[]{yifeng@iphy.ac.cn}
\affiliation{Beijing National Laboratory for Condensed Matter Physics,  Institute of Physics, 
Chinese Academy of Science, Beijing 100190, China}
\affiliation{School of Physical Sciences, University of Chinese Academy of Sciences, Beijing 100190, China}
\affiliation{Songshan Lake Materials Laboratory, Dongguan, Guangdong 523808, China}

\maketitle

\subsection{I. Self-consistent equations}
\begin{figure}[b]
\centering\includegraphics[scale=0.4]{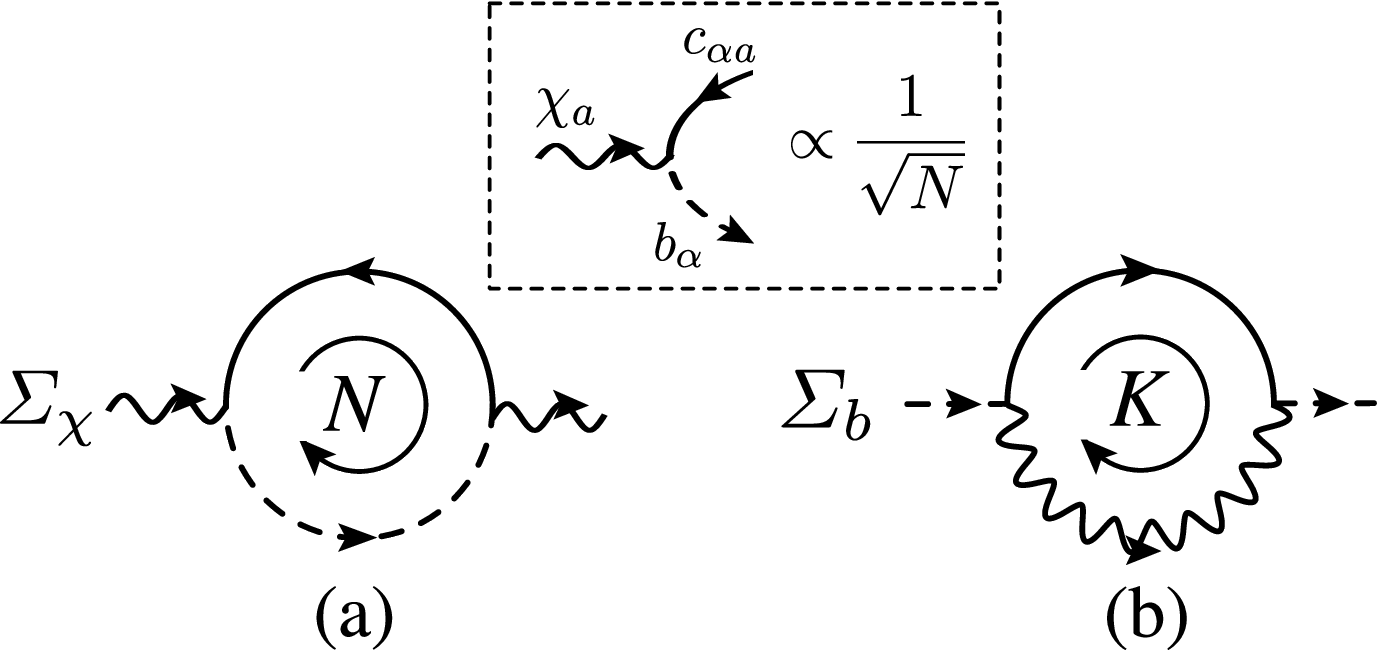}
\caption{The leading order skeleton diagram of (a) $\Sigma_\chi$ and (b) $\Sigma_b$. Circles denote summation over $N$ spin or $K$ channel indices, which cancel the $1/N$ factor from the two vertices. The bare three-point vertex is shown on top and proportional to $1/\sqrt{N}$.}
\label{fig:FeynmD}
\end{figure}

We start with the action of the Schwinger boson Sp($N$) Kondo-Heisenberg model,
\begin{eqnarray}
S &=&-\frac{1}{\beta\mathcal{V}}\sum_{\mathbf{p}m \alpha a}c _{\mathbf{p}m \alpha a}^{\ast }\left( i\omega_m 
-\varepsilon _{\mathbf{p}}\right) c _{\mathbf{p}m \alpha a}-\frac{1}{\beta\mathcal{V}}\sum_{\mathbf{p}n \alpha }b_{\mathbf{p}n \alpha }^{\ast }\left( i\nu_n -\lambda \right) b_{\mathbf{p}n \alpha }  \notag \\
&&+\frac{1}{(\beta\mathcal{V})^2\sqrt{N}}\sum_{\mathbf{pk} m n \alpha a}\left(b_{\mathbf{p}n \alpha }^{\ast
}c_{\mathbf{k}m \alpha a}\chi_{\mathbf{p-k}, n-m-1, a}+h.c.\right)+\frac{1}{\beta\mathcal{V}}\sum_{\mathbf{p} m a}\frac{\left\vert \chi_{\mathbf{p} m a}\right\vert ^{2}}{J_{K}}  \notag \\
&&-\frac{1}{\beta\mathcal{V}}\sum_{\mathbf{p} n \alpha }\left(i\Delta^*\tilde{\alpha}b_{\mathbf{p} n \alpha
}b_{\mathbf{-p}, -n,-\alpha }\xi _{\mathbf{p}}+h.c.\right)+N\beta \mathcal{V}\left( 2\left\vert \Delta \right\vert^{2}/J_{H}-\lambda \kappa \right) ,  \label{action}
\end{eqnarray}
from which the spinon and holon Green's functions can be derived using the Luttinger-Ward functional or the Dyson-Schwinger equations,
\begin{eqnarray} 
G_{b}\left(\mathbf{p}, i\nu_n \right) &\equiv&-\frac{1}{\beta\mathcal{V}}\left\langle b_{\mathbf{p}n \alpha} b_{\mathbf{p}n \alpha}^*\right\rangle =\frac{-\gamma _{b}\left(-\mathbf{p}, -i\nu_n \right) }{4\left\vert \Delta \right\vert ^{2}\xi_{\mathbf{p}}^{2}-\gamma _{b}\left( \mathbf{p},i\nu_n \right) \gamma _{b}\left(-\mathbf{p}, -i\nu_n \right) }, \label{SC4-a} \\
G_{\chi }\left( \mathbf{p},i\omega_m  \right) &\equiv&-\frac{1}{\beta\mathcal{V}}\left\langle \chi_{\mathbf{p}m a} \chi_{\mathbf{p}m a}^*\right\rangle=\frac{1}{-1/J_{K}-\Sigma _{\chi}\left( \mathbf{p},i\omega_m  \right) }, \label{SC4-b} 
\end{eqnarray}
where $\xi _{\mathbf{p}}=\sin p_{x}+\sin p_{y}$ comes from the Fourier transform of the spinon pairing term, and $\gamma_b(\mathbf{p},i\nu_n)\equiv i\nu_n-\lambda-\Sigma_b(\mathbf{p},i\nu_n)$ is the inverse spinon Green's function at $\Delta=0$. The self-energies depend self-consistently on the Green's functions as
\begin{eqnarray}
\Sigma _{b}({\textbf p}, i\nu_n)&=&-\frac{\kappa}{\beta\mathcal{V}} \sum_{{\textbf k}m}g_c({\textbf p}-{\textbf k}, i\nu_n-i\omega_m ) G_{\chi }({\textbf k}, i\omega_m ), \label{SE1} \\
\Sigma _{\chi }({\textbf p}, i\omega_m)&=&\frac{1}{\beta\mathcal{V}}\sum_{{\textbf k}n}g_c({\textbf k}-{\textbf p}, i\nu_n -i\omega_m ) G_{b}({\textbf k}, i\nu_n ), \label{SE2}
\end{eqnarray}
with $g_c(\mathbf{k},i\omega_m )=[i\omega_m +(\cos k_x+\cos k_y)/2]^{-1}$ being the bare Green's function of conduction electrons. The structure of the self-energies can be illustrated via the skeleton Feynman diagrams in Fig.~\ref{fig:FeynmD}. There are no vertex corrections and conduction electron self-energy in the large-$N$ limit.

The variational parameters $\lambda$ and $\Delta$ can be determined via
\begin{equation}
\frac{\partial \ln Z}{\partial \lambda}=\frac{\partial \ln Z}{\partial \Delta}=0,
\end{equation}
where $Z=\int_{[c,b,\chi]}\exp [-S]$ is the partition function of Eq. (\ref{action}). These lead to the following constraints:
\begin{eqnarray}
\kappa &=&-\frac{1}{\beta\mathcal{V}}\sum_{\mathbf{p}n }G_{b}\left(\mathbf{p}, i\nu_n \right), \label{con:1} \\
\frac{1}{J_{H}}\left(1+\frac{12\zeta|\Delta|^2}{J_H^2} \right) &=&-\frac{1}{\beta\mathcal{V}}\sum_{\mathbf{p}n }\frac{\xi
_{\mathbf{p}}^{2}}{4\left\vert \Delta \right\vert ^{2}\xi _{\mathbf{p}}^{2}-\gamma _{b}\left(\mathbf{p},i\nu_n \right) \gamma _{b}\left( \mathbf{-p},-i\nu_n \right) },  \label{con:2}
\end{eqnarray}
where $12\zeta|\Delta|^2/J_H^2$ comes from the biquadratic term, $-\zeta J_H\sum_{\left\langle ij\right\rangle}\left( S_{i}\cdot S_{j} \right)^2$, which is introduced to remedy artificial first-order transitions in the Schwinger boson mean-field theory. In the physical SU(2) case, it can be absorbed into the quadratic exchange term,
\begin{eqnarray}
H_H=J_H\sum_{\left\langle ij\right\rangle}\left(S_{i}\cdot S_{j}-\zeta \left(S_{i}\cdot S_{j}\right)^2\right)=\left(1+\frac{\zeta}{2}\right)J_H\sum_{\left\langle ij\right\rangle}S_{i}\cdot S_{j}+C,
\end{eqnarray}
where $C$ is a constant. In the large-$N$ limit, both terms can be factorized using the mean-field variable $\Delta_{ij}$: 
\begin{eqnarray}
-\frac{J_H}{N}\sum_{\left\langle ij\right\rangle}B_{ij}^\dagger B_{ij}&\rightarrow& \sum_{\left\langle ij\right\rangle}\left(\frac{N}{J_H}|\Delta_{ij}|^2+ B_{ij}^\dagger \Delta_{ij}+\Delta_{ij}^* B_{ij}\right), \label{H_H:1} \\
\frac{2J_H'}{N^3}\sum_{\left\langle ij\right\rangle}\left( B_{ij}^\dagger B_{ij}\right)^2 &\rightarrow& \sum_{\left\langle ij\right\rangle}\left(-\frac{6 N J_H'}{J_H^4}|\Delta_{ij}|^4-\frac{4J_H'}{J_H^3}|\Delta_{ij}|^2\left( B_{ij}^\dagger \Delta_{ij}+\Delta_{ij}^*B_{ij}\right)\right), \label{H_H:2}
\end{eqnarray}
where $B_{ij}\equiv\sum_{\alpha}\tilde{\alpha}b_{i\alpha}b_{j,-\alpha}$, and $J_H'=-\zeta J_H$ has been scaled to $(2/N)^3 J_H'$ in order for a mean-field expansion in terms of $1/N$. The second term on the right hand side of Eq. (\ref{H_H:2}) can be ignored for $-4J_H'|\Delta_{ij}|^2/J_H^3\ll 1$. In our case, a small $\zeta\approx 0.33$ is enough to get rid of all artificial first-order transitions. We find $-4J_H'|\Delta_{ij}|^2/J_H^3\sim 10^{-2}$ near the AFM QCP.

\subsection{II. Two limits: $\kappa=0$ and $J_K=0$}

Based on Eq. (\ref{SE1}) and the fact $G_\chi=0$ at $J_K=0$, it is evident that the spinon self-energy vanishes in both limits. The constraints are then simplified to
\begin{eqnarray}
\kappa &=& \frac{1}{2\mathcal{V}}\sum_\mathbf{p}\frac{\lambda }{\epsilon_\mathbf{p}}\coth\left( \frac{\beta\epsilon_\mathbf{p}}{2}\right)-\frac{1}{2},  \label{ct1} \\ 
\frac{1}{J_H}&=&\frac{1}{2\mathcal{V}}\sum_\mathbf{p}\frac{\xi_\mathbf{p}^2 }{\epsilon_\mathbf{p}}\coth\left( \frac{\beta\epsilon_\mathbf{p}}{2}\right), \label{ct2}
\end{eqnarray} 
where $\epsilon_\mathbf{p}=\sqrt{\lambda^2-4|\Delta|^2\xi_\mathbf{p}^2}$ is the dispersion of free spinons.
 
For $\kappa=0$ and zero temperature ($\beta=\infty$), the above equations require $\Delta=0$ for any finite $J_H$, which implies completely local and independent spinons. We have thus the single impurity Kondo effect in the $\kappa=0$ limit. To show the Fermi surface jump, we note that, because $G_b$ is momentum independent in this limit, the holon self-energy also becomes momentum independent and reduces to
\begin{equation}
\Sigma_\chi(i\omega_m)=\frac{1}{\beta}\sum_{n}g_c(i\nu_n-i\omega_m)G_b(i\nu_n),
\end{equation}
where $g_c(i\omega_m)=\frac{1}{\mathcal{V}}\sum_\textbf{k}g_c(\textbf{k},i\omega_m)$ is the momentum averaged conduction electron Green's function. At zero temperature, the real part of holon self-energy at the Fermi energy can be calculated as
\begin{equation}
\Sigma_\chi'(0)=-\int_{-\infty}^0\frac{dz}{\pi}\frac{g_c''(z)}{z-\lambda},
\end{equation}
which gives rise to a holon Fermi surface jump at some critical value $\lambda=\lambda_c$ satisfying 
\begin{equation}
V_{FS}^\chi=\frac{1}{\mathcal{V}}\sum_\textbf{k}\theta(-J_K^{-1}-\Sigma_\chi'(0))=\begin{cases}
1, & \lambda<\lambda_{c}\\
0, & \lambda>\lambda_{c}
\end{cases}.
\end{equation}
For our model with the chosen conduction band, we find $\lambda_c\approx 0.026$, in good agreement with its numerical value obtained self-consistently as we approach $\kappa=0$ along the HFL boundary.

To see the AFM phase transition in the $J_K=0$ limit, we first introduce the fictitious ``density of states'',
\begin{equation}
\rho_0(z)\equiv \frac{1}{\mathcal{V}}\sum_\mathbf{p}\delta(z-\xi_\mathbf{p}^2)=-\frac{2}{\pi^{2}}\text{Im}\left\{ \frac{1}{z+i0^+}E_{K}\left(\frac{4}{z+i0^+}\right)\right\},
\end{equation}
where $E_K(y)=\int_0^{\pi/2}dx/\sqrt{1-y\sin^2 x}$ is the elliptic integral of the first kind. The constraints become
\begin{eqnarray}
2\kappa+1&=&\int_{0}^4dz\rho_{0}(z)\frac{\lambda}{\sqrt{\lambda^{2}-4|\Delta|^{2}z}}\coth\left(\frac{\beta\sqrt{\lambda^{2}-4|\Delta|^{2}z}}{2}\right),\label{ct21}\\
\frac{2}{J_{H}}&=&\int_{0}^4dz\rho_{0}(z)\frac{z}{\sqrt{\lambda^{2}-4|\Delta|^{2}z}}\coth\left(\frac{\beta\sqrt{\lambda^{2}-4|\Delta|^{2}z}}{2}\right). \label{ct22}
\end{eqnarray}
Defining $Y\equiv 4|\Delta|^2/\lambda^2$, Eq. (\ref{ct21}) can be rewritten as
\begin{equation}
2\kappa+1=\int_{0}^{4}dz\frac{\rho_{0}(z)}{\sqrt{1-Yz}}\coth\left(\frac{\beta \lambda\sqrt{1-Yz}}{2}\right).
\end{equation}
At zero temperature, we have
\begin{equation}
2\kappa+1=\int_{0}^{4}dz\frac{\rho_{0}(z)}{\sqrt{1-Yz}}\leq 1.39.
\end{equation}
Thus the constraint can only be satisfied for $\kappa\le0.195$. The critical $\kappa_c\approx0.195$ occurs when $Y=1/4$, which implies a zero spinon gap at $\textbf{p}=(\pi/2,\pi/2)$. Beyond $\kappa_c$, the ground state is AFM with spinon condensate.

\subsection{III. Numerical algorithm}
Direct calculations of the full self-consistent equations are quite difficult. To reduce computational costs, we develop a  numerical algorithm based on the fast Fourier transform (FFT) for 
\begin{eqnarray}
G(x,y;\omega)=\frac{1}{\mathcal{V}}\sum_{nm}G(n,m;\omega)e^{i2\pi(nx+my)/L},
 \label{FFT} 
\end{eqnarray}
where $\mathbf{r}=(x,y)$ and $\mathbf{k}=2\pi(n,m)/L$ with $x,y,n,m=0,1,\cdots, L-1$. We have $\mathcal{V}=L^2$ and $L$ is the lattice size along one dimension (chosen to be $64$ in our calculations). Equations (\ref{SE1}) and (\ref{SE2}) in the coordinate and real frequency space become 
\begin{eqnarray}
\Sigma_b(\mathbf{r},\omega) &=& -\kappa\int\frac{dz}{\pi}\left[n_F(-z)g_c(\mathbf{r},\omega-z)G_\chi ''(\mathbf{r},z)-n_F(\omega-z)g_c''(\mathbf{r},\omega-z)G_\chi(\mathbf{r},z) \right], \label{SS1} \\
\Sigma_\chi(\mathbf{r},\omega) &=& \int\frac{dz}{\pi}\left[ n_B(z)g_c(-\mathbf{r},z-\omega)^*G_b''(\mathbf{r},z)-n_F(z-\omega)g_c''(-\mathbf{r},z-\omega)G_b(\mathbf{r},z)\right], \label{SS2} 
\end{eqnarray}
where $n_B(z)=1/(e^{\beta z}-1)$ and $n_F(z)=1/(e^{\beta z}+1)$ are the bosonic and fermionic distribution functions, respectively. In each iteration, we first use FFT to transform the Green's functions to the coordinate space, integrating out the above equations at each $\mathbf{r}$, and then transform the self-energies back to the momentum space to calculate the Green's functions. The lattice symmetry has also been taken into consideration such that the Brillouin zone is divided into eight equivalent regions to save the computation time.

\subsection{IV. Evolution of the electron Fermi surface}
\begin{figure}[h]
\centering\includegraphics[scale=0.45]{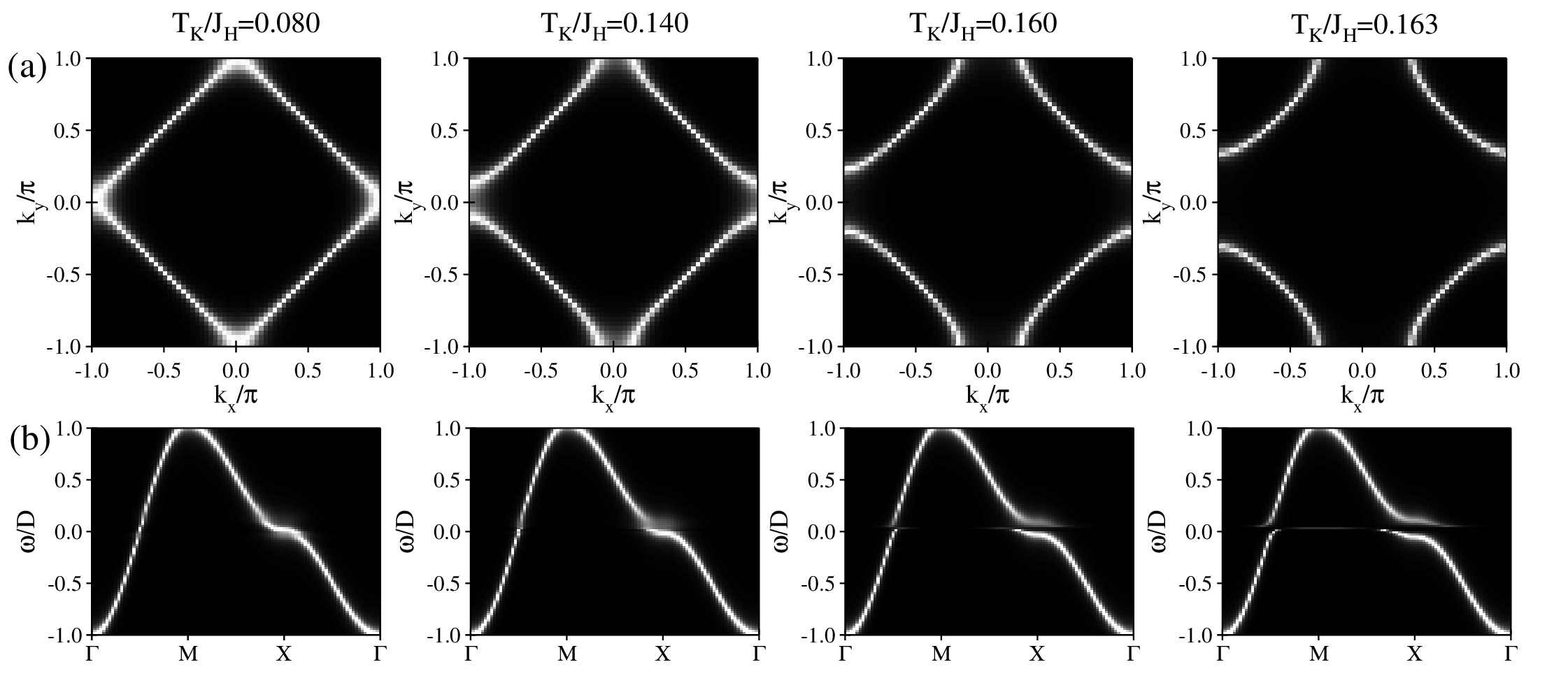}
\caption{(a) Evolution of the electron Fermi surface from the density plot of $A_c(\mathbf{k},\omega=0)$ for $\kappa=0.1$ and $N=6$ at zero temperature. (b) Evolution of the corresponding dispersion from the density plot of $A_c(\mathbf{k},\omega)$ along a high symmetry line in the momentum space for the same parameters. }
\label{fig:EFS1}
\end{figure}

The ``partial" enlargement of the electron Fermi surface in the HS following the generalized Luttinger sum rule may also be seen directly from its spectral function, $A_c(\mathbf{k},\omega)=-\frac{1}{\pi}\text{Im}[1/(\omega -\varepsilon_\mathbf{k}-\Sigma_c(\mathbf{k},\omega))]$, by taking into account the $1/N$ correction to its self-energy, $\Sigma_c(\mathbf{k},i\omega_m )=\frac{1}{\beta\mathcal{V}N}\sum_{\mathbf{p}n}G_\chi(\mathbf{p}-\mathbf{k},i\nu_n-i\omega_m )G_b(\mathbf{p},i\nu_n)$, where $G_b$ and $G_\chi$ are from the infinite $N$ limit. The results are plotted in Fig.~\ref{fig:EFS1} for $\kappa=0.1$ and $N=6$. The electron Fermi surface is gradually enlarged as $T_K/J_H$ increases. Correspondingly, a ``hybridization gap'' is clearly seen to develop near the Fermi energy, starting at around $T_K/J_H=0.14$ in the HS and becomeing fully opened near the HS-to-HFL transition. It is important to note that there is no flat $f$ electron band as in the slave boson theory. Yet the ``hybridization gap" still opens due to the vertex shown in Fig.~\ref{fig:FeynmD}, which produces a sharp resonance peak in the electron self-energy and hence the ``band bending".

\subsection{V. Two-particle correlation functions}
The correlation functions were first calculated in imaginary time,
\begin{eqnarray}
C_O(\mathbf{r}_i-\mathbf{r}_j, \tau-\tau')=-\left\langle O_i(\tau)O_j(\tau') \right\rangle_c,
 \label{sus1} 
\end{eqnarray}
from which the spectra in real frequency can be obtained via analytic continuation. The holon density-density correlation function with $O_i(\tau)=\frac{1}{\sqrt{K}}\sum_a|\chi_{ia}(\tau)|^2$ is given by
\begin{eqnarray}
C_{n_\chi}(\mathbf{p}, i\nu_n)=\frac{1}{\beta\mathcal{V}}\sum_{\mathbf{k}m}G_\chi(\mathbf{k},i\omega_m )G_\chi(\mathbf{k}-\mathbf{p},i\omega_m -i\nu_n)+O(1/N),
 \label{chargesus} 
 \end{eqnarray}
 where the $O(1/N)$ term contains four-point connected diagrams that can be ignored in the large-$N$ limit. The spin correlation function is defined with $O_i(\tau)=\frac{1}{\sqrt{N}}\sum_\alpha \tilde{\alpha}b_{i\alpha}^\dagger (\tau)b_{i\alpha}(\tau)$. Outside the AFM phase, it reduces to
\begin{eqnarray}
C_{S}(\mathbf{p}, i\nu_n)=-\frac{1}{\beta\mathcal{V}}\sum_{\mathbf{k}n'}\left[G_b(\mathbf{k},i\nu_{n'})G_b(\mathbf{k}+\mathbf{p},i\nu_{n'}+i\nu_n)-F_b(\mathbf{k},i\nu_{n'})\bar{F}_b(\mathbf{k}+\mathbf{p},i\nu_{n'}+i\nu_n)\right]+O(1/N),
 \label{spinsus} 
 \end{eqnarray}
 where $F_{b}\left(\mathbf{p}, i\nu_n \right)=-(1/\beta\mathcal{V})\tilde{\alpha}\left\langle b_{\mathbf{p}n\alpha} b_{-\mathbf{p},-n,-\alpha}   \right\rangle$ is the anomalous Green's function due to the finite amplitude of spinon pairing. The staggered spin susceptibility is given by $\chi_{st}=-C_{S}(\mathbf{Q}, 0)$ at $\textbf{Q}=(\pi,\pi)$.

\begin{figure}[h]
\centering\includegraphics[scale=0.5]{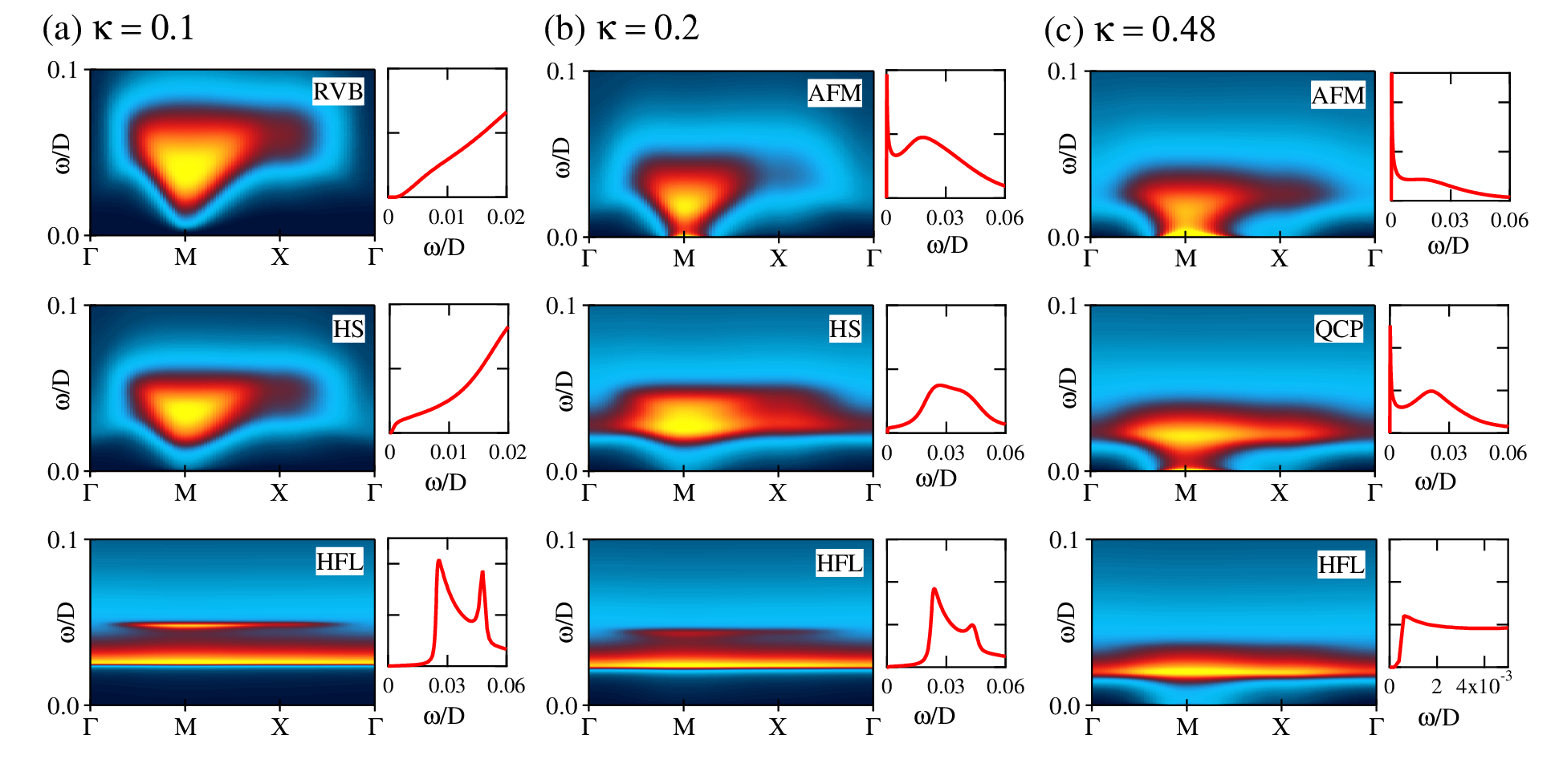}
\caption{The spin spectral function, $A_S(\mathbf{k},\omega)=-C_S''(\mathbf{k},\omega)/\pi$, and their profiles at the $M$ point above different ground states (or QCP) for (a) $\kappa=0.1$, (b) $\kappa=0.2$ and (c) $\kappa=0.48$. The calculations were performed in the paramagnetic phase at a small but finite temperature. The label ``AFM" is only an indication of the AFM ground state at zero temperature.}
\label{fig:spinspec}
\end{figure}

Figure \ref{fig:spinspec} plots the spin spectral function, $A_S(\mathbf{k},\omega)=-C_S''(\mathbf{k},\omega)/\pi$, for different values of $T_K/J_H$ for $\kappa=0.1$, $0.2$ and $0.48$ at a small but finite temperature. We see it develops a sharp peak at the $M$ point at $\omega=0$ above the AFM ground state or at the AFM QCP, which we ascribe to the precursor of spinon condensation. By contrast, $A_S(\mathbf{k},\omega)$ shows a gap at $\omega=0$ in the HFL phase, indicating the confinement of spinons and gauge fields for large Kondo coupling. In the intermediate HS region, both features are missing and we find gapless and damped spinon excitations.
 
 \subsection{VI. Magnetic entropy}
The magnetic entropy (divided by $N$) has the following expression \cite{coleman2005sum},
\begin{eqnarray}
S&=&- \frac{1}{\mathcal{V}}\sum_\mathbf{k}\int\frac{dz}{\pi}\left\lbrace  \frac{dn_B}{dT}\left(\frac{1}{2}\text{Im}\ln \left[\gamma_b(\mathbf{k},z) \gamma_b(-\mathbf{k},-z)^*-4|\Delta|^2\xi_\mathbf{k}^2\right]+\Sigma_b''(\mathbf{k},z)G_b'(\mathbf{k},z)   \right) \right.  \notag \\
 & &\qquad +\left. \kappa \frac{dn_F}{dT}\left( \text{Im}\ln \left[ -G_\chi^{-1}(\mathbf{k},z) \right] +\Sigma_\chi''(\mathbf{k},z)G_\chi'(\mathbf{k},z)-N\Sigma_c'(\mathbf{k},z) g_c''(\mathbf{k},z)   \right)\right\rbrace,
 \label{entropy} 
 \end{eqnarray}
where $\Sigma_c(\mathbf{k},i\omega_m )=\frac{1}{\beta\mathcal{V}N}\sum_{\mathbf{p}n}G_\chi(\mathbf{p}-\mathbf{k},i\nu_n-i\omega_m )G_b(\mathbf{p},i\nu_n)$ is the conduction electron self-energy. The free electron contribution has been excluded. The specific heat coefficient can be calculated using $C/T=dS/dT$.